\documentclass[12pt]{article}
\usepackage{a4}
\usepackage{amsmath}
\usepackage{amsfonts}
\usepackage{amsmath}
\usepackage{latexsym}
\usepackage{amsfonts}
\begin{document}
\oddsidemargin 6pt\evensidemargin 6pt\marginparwidth
48pt\marginparsep 10pt

\renewcommand{\thefootnote}{\fnsymbol{footnote}}
\thispagestyle{empty}

\noindent    \hfill  \\

\noindent \vskip3.3cm
\begin{center}

{\Large\bf Conformal invariant interaction  of a scalar field with the higher spin field in
$AdS_{D}$}
\bigskip\bigskip\bigskip

{\large
Ruben Manvelyan and Karapet Mkrtchyan}\\
\medskip
{\small\it Yerevan Physics Institute\\ Alikhanian Br.
Str.
2, 0036 Yerevan, Armenia}\\
\medskip
{\small\tt e-mails manvel@physik.uni-kl.de, karapet@yerphi.am}
\end{center}

\bigskip 
\begin{center}
{\sc Abstract}
\end{center}

The explicit form of linearized gauge invariant interactions of scalar and general higher even spin fields in the $AdS_{D}$ space is obtained.
In the case of general spin $\ell$  a generalized 'Weyl' transformation is proposed and the corresponding 'Weyl' invariant action is constructed.
In both cases the invariant actions of the interacting higher even spin gauge field and the scalar field include the whole tower of invariant actions for couplings of the same scalar with all gauge fields of smaller even spin. For the particular value of $\ell=4$ all results are in exact agreement with \cite{WI}

\newpage

\section{Introduction}
\quad
After discovering the $AdS_{4}/CFT_{3}$
correspondence of the critical $O(N)$ sigma model \cite{Klebanov},
interest in the interacting theory of an arbitrary even high spin field drastically increased.
So in the center of our attention is a theory of Fradkin-Vasiliev type \cite{Vasiliev}-\cite{Vasiliev3} in Fronsdal's metric formulation
\cite{Fronsdal}, \cite{Fronsdal1}. This case of $AdS_{D}/CFT_{D-1}$ correspondence is also of great interest because on the one hand supersymmetry and BPS arguments are absent and on the other hand  both conformal points of the boundary theory (i.e. unstable free field
theory and critical interacting theory, in the large $N$ limit)
correspond to the same higher spin theory, and are connected on the
boundary by a Legendre transformation which corresponds to
different boundary conditions (regular dimension one or
shadow dimension two) in the quantization of the bulk scalar field \cite{WittKleb}-\cite{WittKleb2}.
Existence of this scalar field  in higher spin gauge theory is also an interesting and
important phenomenon and supports the spontaneous symmetry breaking mechanism and mass
creation for initially  massless gauge fields due to corresponding possible interactions (see for example \cite{MMR}-\cite{MR22}).

From this point of view any construction of a reasonable even linearized interaction is an interesting and important task in this reconstruction of the higher spin gauge theory from the holographic dual CFT and can be controlled by the corresponding information about the anomalous dimensions of the dual global symmetry currents that fulfill  the conservation conditions in the large $N$ limit.
Therefore we see that a construction of the conformal coupling of the scalar with a general even higher spin gauge field appears as an interesting  example of an interaction which is applicable to many different quantum one-loop calculations such as the trace anomaly of the scalar in the external higher spin gauge field and so on \cite{MRA}-\cite{MRA2}.

In this article we construct a generalization of the well known action for the conformally coupled scalar
field in $D$ dimensions in external gravity
\begin{equation}\label{1}
    S=\frac{1}{2}\int
    d^{D}z\sqrt{-G}\left[G^{\mu\nu}\nabla_{\mu}\phi\nabla_{\nu}\phi
    -\frac{(D-2)}{4(D-1)}R(G)\phi^{2}\right] .
\end{equation}
to the coupling with the linearized external higher spin $\ell$ gauge field.
Actually in this article we accomplished to generalize the result of \cite{WI} for
spin four, obtained four years ago, to the general spin $\ell$ case.
We show that the gauge  and 'Weyl' invariant interaction of the
scalar with the spin $\ell$ Fronsdal gauge field can be constructed only if we  add
the same type of interaction with all lower even spin gauge fields. In other words we can
construct a self-consistent interaction of a gauge field with the conformally coupled scalar only with the whole
finite tower of gauge fields with spins in the range $2\leq s\leq \ell$.
In the next section we fix the notation and conventions and briefly review the results of \cite{WI}.
In section 3 we explicitly construct a linearized interaction \emph{Lagrangian} of
the conformal scalar field with the spin $\ell$ gauge field
using Noether's procedure for higher spin \emph{gauge} invariance. In section 4 we extend our investigation including Noether's procedure for \emph{generalized
Weyl invariance} and obtain a unique interacting action after nontrivial and tedious calculations summarized in several appendices.
Note also that nonlinear gauge invariant
couplings of the scalar field on the level of the equation of motion
were under consideration in \cite{ss}, \cite{ss1}  and on the level of the BRST formalism for higher spin fields in \cite{Petkou}. Summarizing the introduction we can
say that this is a linearized interaction with the scalar field for \emph{conformal
higher spin theory} of the type discussed by \cite{Segal}, \cite{Fradkin}-\cite{Fradkin2}.

\section{The cases of spin two and spin four}

\quad We work in Euclidian $AdS_{D}$ with the following metric,
curvature and covariant derivatives:
\begin{eqnarray}
&&ds^{2}=g_{\mu \nu }(z)dz^{\mu }dz^{\nu
}=\frac{L^{2}}{(z^{0})^{2}}\delta _{\mu \nu }dz^{\mu }dz^{\nu
},\quad \sqrt{-g}=\frac{L^{D}}{(z^{0})^{D}}\;,
\notag  \label{ads} \\
&&\left[ \nabla _{\mu },\,\nabla _{\nu }\right] V_{\lambda }^{\rho }=R_{\mu
\nu \lambda }^{\quad \,\,\sigma }V_{\sigma }^{\rho }-R_{\mu \nu \sigma
}^{\quad \,\,\rho }V_{\lambda }^{\sigma }\;,  \notag \\
&&R_{\mu \nu \lambda }^{\quad \,\,\rho
}=-\frac{1}{(z^{0})^{2}}\left( \delta _{\mu \lambda }\delta _{\nu
}^{\rho }-\delta _{\nu \lambda }\delta _{\mu }^{\rho }\right)
=-\frac{1}{L^{2}}\left( g_{\mu \lambda }(z)\delta _{\nu
}^{\rho }-g_{\nu \lambda }(z)\delta _{\mu }^{\rho }\right) \;,  \notag \\
&&R_{\mu \nu }=-\frac{D-1}{(z^{0})^{2}}\delta _{\mu \nu }=-\frac{D-1}{L^{2}}%
g_{\mu \nu }(z)\quad ,\quad R=-\frac{(D-1)D}{L^{2}}\;.  \notag
\end{eqnarray}%

In \cite{WI} the authors constructed gauge and generalized Weyl invariant actions for spin two and four gauge fields
interacting with a scalar field. Here we review these results in the form suitable for a generalization to arbitrary higher even spin fields. We work with double traceless higher spin fields in Fronsdal's formulation \cite{Fronsdal},\cite{Fronsdal1} where the free field equation of motion for the higher spin $\ell$ field $h_{\mu_{1}...\mu_{s}}$ reads

\begin{eqnarray}
\mathcal{F}_{\mu_{1}...\mu_{\ell}}=\Box h_{\mu_{1}...\mu_{\ell}}
-\ell \nabla_{(\mu_{1}}\nabla^{\rho}h_{\mu_{2}...\mu_{\ell})\rho}
+\frac{\ell(\ell-1)}{2}\nabla_{(\mu_{1}}\nabla_{\mu_{2}}h^{\ \ \ \ \ \ \ \ \rho}_{\mu_{3}...\mu_{\ell})\rho}\nonumber\\
+\frac{\ell^{2}+\ell(D-6)-2(D-3)}{L^{2}}h_{\mu_{1}...\mu_{\ell}}
+\frac{\ell(\ell-1)}{L^{2}}g_{(\mu_{1}\mu_{2}}h^{\ \ \ \ \ \ \ \ \rho}_{\mu_{3}...\mu_{\ell})\rho}=0
\end{eqnarray}

This equation is invariant under gauge transformation\footnote{We denote symmetrization of indices by
round brackets.}

\begin{eqnarray}
\delta h_{\mu_{1}...\mu_{\ell}}=\ell\nabla_{(\mu_{1}}\epsilon_{\mu_{2}...\mu_{\ell})}=\nabla_{\mu_{1}}\epsilon_{\mu_{2}...\mu_{\ell}}+c.p.
\end{eqnarray}
where

\begin{eqnarray}
h^{\ \ \ \ \ \ \ \ \ \ \ \rho\sigma}_{\mu_{1}...\mu_{\ell-4}\rho\sigma}=0,\\
\epsilon^{\ \ \ \ \ \ \ \ \ \ \rho}_{\mu_{1}...\mu_{\ell-3}\rho}=0.
\end{eqnarray}
The trace of Fronsdal's tensor reads as

\begin{eqnarray}
r^{(\ell)\mu_{1}...\mu_{\ell-2}}=-\frac{1}{2}Tr\mathcal{F}(h^{\ell})=\nabla_{\alpha}\nabla_{\beta}h^{(\ell)\alpha\beta\mu_{1}...\mu_{\ell-2}}-\Box h_{\alpha}^{(\ell)\alpha\mu_{1}...\mu_{\ell-2}}\nonumber\\
-\frac{\ell-2}{2}\nabla^{(\mu_{1}}\nabla_{\alpha}h_{\beta}^{(\ell)\mu_{2}...\mu_{\ell-2})\alpha\beta}
-\frac{(\ell-1)(D+\ell-3)}{L^{2}}h_{\alpha}^{(\ell)\alpha\mu_{1}...\mu_{\ell-2}}.\label{frdtr}
\end{eqnarray}

For the case $\ell=2$  one can see \cite{WI} that a Weyl invariant action is
\begin{eqnarray}
&&S^{WI}(\phi,h^{(2)})=S_{0}(\phi)+S_{1}^{\Psi^{(2)}}(\phi,h^{(2)})+S_{1}^{r^{(2)}}(\phi,h^{(2)}). \label{spin2}
\end{eqnarray}
where
\begin{eqnarray}
S_{0}(\phi)&=&\frac{1}{2}\int d^{D}z\sqrt{-g}[\nabla_{\mu}\phi\nabla^{\mu}\phi+\frac{D(D-2)}{4L^{2}}\phi^{2}], \label{S0}\\
S_{1}^{\Psi^{(2)}}(\phi,h^{(2)})&=&\frac{1}{2}\int d^{D}z\sqrt{-g}h^{(2)\mu\nu}\Psi^{(2)}_{\mu\nu}(\phi)\label{psi12}\\\Psi^{(2)}_{\mu\nu}(\phi)&=&-\nabla_{\mu}\phi\nabla_{\nu}\phi
+\frac{g_{\mu\nu}}{2}(\nabla_{\lambda}\phi\nabla^{\lambda}\phi+\frac{D(D-2)}{4L^{2}}\phi^{2}), \quad\label{psi2}\\
S_{1}^{r^{(2)}}(\phi,h^{(2)})&=&\frac{1}{8}\frac{D-2}{D-1}\int d^{D}z\sqrt{-g}r^{(2)}(h^{(2)})\phi^{2}, \label{Sr2}\\r^{(2)}(h^{(2)})&=&\nabla_{\mu}\nabla_{\nu}h^{(2)\mu\nu}-\Box h_{\mu}^{(2)\mu}
-\frac{D-1}{L^{2}}h_{\mu}^{(2)\mu}\label{Sr12}
\end{eqnarray}
which is of course the linearized form of (\ref{1}) and is invariant with respect to the gauge and Weyl transformations
\footnote{$\Delta$ is so-called conformal weight of the scalar and gets fixed by conformal invariance condition}
\begin{eqnarray}
\delta_{\varepsilon}^{1}\phi&=&\varepsilon^{\mu}(z)\nabla_{\mu}\phi\label{phig2},\quad
\delta_{\varepsilon}^{0}h_{\mu\nu}^{(2)}=2\nabla_{(\mu}\varepsilon_{\nu)}\label{eps2};\\
\delta_{\sigma}^{1}\phi(z)&=&\Delta\sigma(z)\phi(z)\label{phiw2},\quad
\delta_{\sigma}^{0}h_{\mu\nu}^{(2)}=2\sigma(z)g_{\mu\nu}\label{hw2}.\\
\Delta&=&1-\frac{D}{2}
\end{eqnarray}

Now we turn to the case $\ell=4$. In \cite{WI} the authors started from the action (\ref{S0}) and applied Noether's procedure using the following higher spin 'reparametrization' of the scalar field with a traceless third rank symmetric tensor parameter
\begin{eqnarray}
\delta_{\epsilon}^{1}\phi(z)=\epsilon^{\mu\nu\lambda}(z)\nabla_{\mu}\nabla_{\nu}\nabla_{\lambda}\phi(z),\quad
\epsilon^{\alpha}_{\alpha\mu}=0.\label{phig4}
\end{eqnarray}
The variation of (\ref{S0}) is\footnote{For compactness  we introduce shortened notations for divergences of the tensor's symmetry parameters
\begin{eqnarray}
\epsilon_{(1)}^{\mu\nu\dots}=\nabla_{\lambda}\epsilon^{\lambda\mu\nu\dots},\quad \epsilon_{(2)}^{\mu\dots}=\nabla_{\nu}\nabla_{\lambda}\epsilon^{\nu\lambda\mu\dots}, \quad\dots
\end{eqnarray}}
\begin{eqnarray}
&&\delta_{\epsilon}^{1}S_{0}(\phi)=\int d^{D}z\sqrt{-g}\{-\nabla^{(\alpha}\epsilon^{\mu\nu\lambda)}\nabla_{\mu}
\nabla_{\alpha}\phi\nabla_{\nu}\nabla_{\lambda}\phi+\epsilon_{(1)}^{\mu\nu}[\frac{1}{2}\nabla_{\mu}
\nabla_{\alpha}\phi\nabla_{\nu}\nabla^{\alpha}\phi\nonumber\\
&&+\frac{D(D+2)}{8L^{2}}\nabla_{\mu}
\phi\nabla_{\nu}\phi]
-\nabla^{(\mu}\epsilon_{(2)}^{\nu)}
[-\nabla_{\mu}\phi\nabla_{\nu}\phi+\frac{g_{\mu\nu}}{2}
(\nabla_{\lambda}\phi\nabla^{\lambda}\phi+\frac{D(D-2)}{4L^{2}}\phi^{2})]\nonumber\\&&+[\nabla^{2}\phi
-\frac{D(D-2)}{4L^{2}}\phi]\nabla_{\mu}(\epsilon_{(1)}^{\mu\nu}\nabla_{\nu}\phi)
 \}.\label{var4}
\end{eqnarray}
We see immediately  that the first two lines of (\ref{var4}) produce interactions with the spin four and two currents. From the other hand the last line in (\ref{var4}) is proportional to the equation of motion following from $S_{0}(\phi)$ and  therefore  can be absorbed after gauging by the trace of the spin four gauge field  ($2\epsilon_{(1)}^{\mu\nu}\rightarrow h_{\alpha}^{(4)\alpha\mu\nu}$) performing the following field redefinition of $\phi$
\begin{equation}\label{frd}
    \phi\rightarrow \phi +\frac{1}{2}\nabla_{\mu}(h_{\alpha}^{(4)\alpha\mu\nu}\nabla_{\nu}\phi)
\end{equation}
Such a type of field redefinition is a standard correction of Noether's procedure and means that we always can drop from the cubic part of the action terms proportional to the equation of motion following from the quadratic part of the initial action.

So finally we see that the action
\begin{eqnarray}
S^{GI}(\phi,h^{(2)},h^{(4)})&=&S_{0}(\phi)+S_{1}^{\Psi^{(2)}}(\phi,h^{(2)})+S_{1}^{\Psi^{(4)}}(\phi,h^{(4)})\label{GI4},
\end{eqnarray}
where $S_{0}(\phi)$ , $S_{1}^{\Psi^{(2)}}(\phi,h^{(2)})$ are defined in (\ref{S0})-(\ref{psi2}) and
\begin{eqnarray}
&&S_{1}^{\Psi^{(4)}}(\phi,h^{(4)})=\frac{1}{4}\int d^{D}z\sqrt{-g}h^{(4)\mu\nu\alpha\beta}\Psi^{(4)}_{\mu\nu\alpha\beta}(\phi)\\
&&\Psi^{(4)}_{\mu\nu\alpha\beta}(\phi)=\nabla_{(\mu}
\nabla_{\nu}\phi\nabla_{\alpha}\nabla_{\beta)}\phi-g_{(\mu\nu}[\nabla_{\alpha}
\nabla^{\gamma}\phi\nabla_{\beta)}\nabla_{\gamma}\phi
+\frac{D(D+2)}{4L^{2}}\nabla_{\alpha}
\phi\nabla_{\beta)}\phi]\label{psi4},\quad\quad
\end{eqnarray}
is invariant with respect to the gauge transformations of the spin four field with  an additional  spin two field gauge transformation inspired by the second divergence of the spin four gauge parameter\footnote{Note that the spin two part of our action continues to be invariant in respect of usual linearized  reparametrization (\ref{eps2})}
\begin{eqnarray}
&&\delta_{\epsilon}^{1}\phi(z)=\epsilon^{\mu\nu\lambda}(z)\nabla_{\mu}\nabla_{\nu}\nabla_{\lambda}\phi(z)
,\label{phig4n}\\
&&\delta_{\epsilon}^{0}h^{(4)\mu\nu\alpha\beta}=4\nabla^{(\mu}\epsilon^{\nu\alpha\beta)},\quad
\delta_{\epsilon}^{0}h_{\alpha}^{(4)\alpha\mu\nu}=2\epsilon_{(1)}^{\mu\nu}\label{eps4},\\
&&\delta_{\epsilon}^{0}h^{(2)\mu\nu}=2\nabla^{(\mu}\epsilon_{(2)}^{\nu)}.\label{eps24}
\end{eqnarray}
 Thus we introduced a gauge invariant interaction of the scalar with the spin four gauge field $h^{(4)}_{\mu\nu\alpha\beta}$ in the minimal way. The next step is the spin four Weyl invariant interaction.

We write the generalized Weyl transformation law for the spin four case  as in the \cite{WI}
\begin{eqnarray}
\delta_{\sigma}^{0}h^{(4)\mu\nu\alpha\beta}(z)=12\sigma^{(\mu\nu}(z)g^{\alpha\beta)},\ \ \
\delta_{\sigma}^{1}\phi(z)=\Delta_{4}\sigma^{\alpha\beta}\nabla_{\alpha}\nabla_{\beta}\phi,\label{sig4}
\end{eqnarray}
where we introduced a generalized "conformal" weight $\Delta_{4}$ for the scalar field.
Then following \cite{WI} one can make (\ref{GI4}) Weyl invariant  introducing the following terms
\begin{eqnarray}
S_{1}^{r^{(4)}}=\frac{1}{2}\xi_{4}^{1}\int d^{D}z\sqrt{-g}r^{(4)\mu\nu}\nabla_{\mu}\phi\nabla_{\nu}\phi+
\frac{1}{2}\xi_{4}^{0}\int d^{D}z\sqrt{-g}\nabla_{\mu}\nabla_{\nu}r^{(4)\mu\nu}\phi^{2},\label{Sr4}
\end{eqnarray}
where\footnote{We have to mention that our $\Delta_{4}$ here differs from $\widetilde{\Delta}$ in \cite{WI} because of field redefinition (\ref{frd}) which is the reason why $S_{1}^{\Psi^{(4)}}$ from \cite{WI} turned into (\ref{psi4}). When we make field redefinition, we add to the Lagrangian terms which are not Weyl invariant, and in order to restore Weyl invariance we have to change the coefficient $\Delta_{4}$.}
\begin{eqnarray}
&&r^{(4)\mu\nu}=\nabla_{\alpha}\nabla_{\beta}h^{(4)\alpha\beta\mu\nu}-\Box h_{\alpha}^{(4)\alpha\mu\nu}-\nabla^{(\mu}\nabla_{\beta}h_{\alpha}^{(4)\nu)\beta\alpha}-\frac{3(D+1)}{L^{2}}h_{\alpha}^{(4)\alpha\mu\nu},\\
&&\delta_{\epsilon}^{1}r^{(4)\mu\nu}=0,\ \ \ \ r_{\mu}^{(4)\mu}=0,\\
&&\xi_{4}^{1}=-\frac{1}{4}\frac{D}{D+3},\ \ \ \ \xi_{4}^{0}=\frac{1}{32}\frac{D(D-2)}{(D+1)(D+3)},\ \ \ \ \ \Delta_{4}=\Delta=1-\frac{D}{2}.
\end{eqnarray}
Thus the linearized action for a scalar field interacting with the spin two and four fields in a conformally invariant way is
\begin{eqnarray}
S^{WI}(\phi,h^{(2)},h^{(4)})=S^{WI}(\phi,h^{(2)})+S_{1}^{\Psi^{(4)}}(\phi,h^{(4)})+S_{1}^{r^{(4)}}(\phi,h^{(4)}),\label{spin4}
\end{eqnarray}
which is invariant with respect to gauge and generalized Weyl transformations
\begin{eqnarray}
&&\delta^{1}\phi=\varepsilon^{\mu}\nabla_{\mu}\phi+\epsilon^{\mu\nu\lambda}\nabla_{\mu}\nabla_{\nu}\nabla_{\lambda}\phi
+\Delta\sigma\phi+\Delta\sigma^{\mu\nu}\nabla_{\mu}\nabla_{\nu}\phi,\\
&&\delta^{0}h^{(2)\mu\nu}=2\nabla^{(\mu}\varepsilon^{\nu)}+2\nabla^{(\mu}\epsilon_{(2)}^{\nu)}
+2(1-\Delta-4D\xi_{4}^{1})\nabla^{(\mu}\sigma_{(1)}^{\nu)}\nonumber\\
&&+2\sigma g^{\mu\nu}+2\xi_{4}^{1}\sigma_{(2)}g^{\mu\nu}\label{rgp}\\
&&\delta^{0}h^{(4)\mu\nu\alpha\beta}=4\nabla^{(\mu}\epsilon^{\nu\alpha\beta)}+12\sigma^{(\mu\nu}g^{\alpha\beta)}.
\end{eqnarray}

\section{Gauge invariant interaction for the spin $\ell$ case}

\quad Here we generalize our construction to the general spin  $\ell$ case.
Again following \cite{WI} we apply the following gauge transformation
\begin{eqnarray}
&&\delta_{\epsilon}^{1}\phi(z)=\epsilon^{\mu_{1}\mu_{2}...\mu_{\ell-1}}(z)\nabla_{\mu_{1}}\nabla_{\mu_{2}}...
\nabla_{\mu_{\ell-1}}\phi(z),\label{phigl}\\
&&\delta_{\epsilon}^{0}h^{(\ell)\mu_{1}...\mu_{\ell}}=l\nabla^{(\mu_{\ell}}\epsilon^{\mu_{1}\mu_{2}...\mu_{\ell-1})},\quad
\delta_{\epsilon}^{0}h_{\alpha}^{(\ell)\alpha\mu_{1}...\mu_{\ell-2}}=2\epsilon_{(1)}^{\mu_{1}...\mu_{\ell-2}}\label{tepsl},\\  &&\epsilon^{\alpha}_{\alpha\mu_{3}...\mu_{\ell-1}}=0
\end{eqnarray}
to the action (\ref{S0}) and obtain  the  following starting variation for Noether's procedure
\begin{eqnarray}
&&\delta_{\epsilon}^{1}S_{0}(\phi)=
\int d^{D}z\sqrt{-g}
\{
\nabla^{\alpha}\epsilon^{\mu_{1}...\mu_{\ell-1}}\nabla_{\alpha}\phi\nabla_{\mu_{1}}...\nabla_{\mu_{\ell-1}}\phi+\nonumber\\
&&\epsilon^{\mu_{1}...\mu_{\ell-1}}\nabla_{\alpha}\phi\nabla^{\alpha}\nabla_{\mu_{1}}...\nabla_{\mu_{\ell-1}}\phi+
\frac{D(D-2)}{4L^{2}}\epsilon^{\mu_{1}...\mu_{\ell-1}}\phi\nabla_{\mu_{1}}...\nabla_{\mu_{\ell-1}}\phi
\}.\label{varl}
\end{eqnarray}
Using the following notations
\begin{eqnarray}
T(n,k)&=&\nabla^{\alpha}\epsilon_{(\ell-n)}^{\mu_{1}...\mu_{n-1}}\nabla_{\mu_{1}}...\nabla_{\mu_{k-1}}\nabla_{\alpha}\phi
                                                               \nabla_{\mu_{k}}...\nabla_{\mu_{n-1}}\phi,\\
M(n,k)&=&\epsilon_{(\ell-n-1)}^{\mu_{1}...\mu_{n}}\nabla_{\mu_{1}}...\nabla_{\mu_{k}}\nabla_{\alpha}\phi
                                              \nabla_{\mu_{k+1}}...\nabla_{\mu_{n}}\nabla^{\alpha}\phi,\\
N(n,k)&=&\epsilon_{(\ell-n-1)}^{\mu_{1}...\mu_{n}}\nabla_{\mu_{1}}...\nabla_{\mu_{k}}\phi
                                              \nabla_{\mu_{k+1}}...\nabla_{\mu_{n}}\phi.
\end{eqnarray}
and commutation relation (B.1) from Appendix B we rewrite (\ref{varl}) in the form
\begin{eqnarray}
&&\delta_{\epsilon}^{1}S_{0}(\phi)=
\int d^{D}z\sqrt{-g}\{T(\ell,1)+M(\ell-1,0)+\nonumber\\
&&+\frac{(\ell-1)(\ell-2)}{2L^{2}}N(\ell-1,1)+\frac{D(D-2)}{4L^{2}}N(\ell-1,0)
\}.\label{varl1}
\end{eqnarray}
Then using relations between $T(m,n)$, $M(m,n)$ and $N(m,n)$ from Appendix A  and after some algebra we 'diagonalize' (\ref{varl1})
\begin{eqnarray}
&&\delta_{\epsilon}^{1}S_{0}(\phi)=\sum_{m=1}^{\frac{\ell}{2}}(-1)^{m}\binom{\ell-m-1}{m-1}
\int d^{D}z\sqrt{-g}\{
-T(2m,m)+\frac{1}{2}M(2m-2,m-1)\nonumber\\
&&+\frac{(D+2m-2)(D+2m-4)}{8L^{2}}N(2m-2,m-1)\nonumber\\
&&-\frac{m-1}{\ell-2m+1}\epsilon_{(\ell-2m+1)}^{\mu_{1}...\mu_{2m-2}}
(\nabla_{\mu_{1}}...\nabla_{\mu_{m-1}}[\nabla^{2}\phi-\frac{D(D-2)}{4L^{2}}\phi]
\nabla_{\mu_{m}}...\nabla_{\mu_{2m-2}}\phi)\}\label{epslS}
\end{eqnarray}
Further performing a final symmetrization in (\ref{epslS}), we obtain the following elegant expression
\begin{eqnarray}
&&\delta_{\epsilon}^{1}S_{0}(\phi)=\int d^{D}z\sqrt{-g}\Big\{\sum_{m=1}^{\frac{\ell}{2}}\binom{\ell-m-1}{m-1}
[-\nabla^{(\mu_{2m}}\epsilon_{(\ell-2m)}^{\mu_{1}...\mu_{2m-1})}\Psi^{(2m)}_{\mu_{1}...\mu_{2m}}]\nonumber\\
&&+[\nabla^{2}\phi-\frac{D(D-2)}{4L^{2}}\phi]\sum_{m=2}^{\frac{\ell}{2}}\binom{\ell-m-1}{m-2}
\nabla_{\mu_{1}}...\nabla_{\mu_{m-1}}(\epsilon_{(\ell-2m+1)}^{\mu_{1}...\mu_{2m-2}}
\nabla_{\mu_{m}}...\nabla_{\mu_{2m-2}}\phi)
\Big\},\nonumber\\\label{varlf}
\end{eqnarray}
where
\begin{eqnarray}
&&\Psi^{(2m)}_{\mu_{1}...\mu_{2m}}=(-1)^{m}\{\nabla_{\mu_{1}}...\nabla_{\mu_{m}}\phi
\nabla_{\mu_{m+1}}...\nabla_{\mu_{2m}}\phi\nonumber\\&&-\frac{m}{2}g_{\mu_{2m-1}\mu_{2m}}g^{\alpha\beta}
\nabla_{(\mu_{1}}...\nabla_{\mu_{m-1}}\nabla_{\alpha)}\phi
\nabla_{(\mu_{m}}...\nabla_{\mu_{2m-2}}\nabla_{\beta)}\phi\nonumber\\
&&-\frac{m(D+2m-2)(D+2m-4)}{8L^{2}}g_{\mu_{2m-1}\mu_{2m}}\nabla_{\mu_{1}}...\nabla_{\mu_{m-1}}\phi
\nabla_{\mu_{m}}...\nabla_{\mu_{2m-2}}\phi\}\label{psil}
\end{eqnarray}
and we admitted symmetrization for the set $\mu_{1},\dots \mu_{2m}$ of indices.
So we see that miraculously the coefficients in (\ref{psil}) don't depend on $\ell$ ! All $\ell$- dependence is concentrated in the second line of (\ref{varlf}) proportional to the equation of motion for the action (\ref{S0}). This part like in the spin four case can be removed by an appropriate field redefinition (see (\ref{tepsm}), (\ref{vareps}), (B.6))
\begin{eqnarray}
    \phi\rightarrow \phi +\sum_{m=2}^{\frac{\ell}{2}}\frac{m-1}{2(\ell-2m+1)}\nabla_{\mu_{1}}...\nabla_{\mu_{m-1}}(h_{\alpha}^{(2m)\alpha\mu_{1}...\mu_{2m-2}}
                                                                                        \nabla_{\mu_{m}}...\nabla_{\mu_{2m-2}}\phi)
\end{eqnarray}
and we can drop these terms from our consideration.
Thus we arrive at the following spin $\ell$ gauge invariant action
\begin{eqnarray}
&&S^{GI}(\phi,h^{(2)},h^{(4)},...,h^{(\ell)})=S_{0}(\phi)+\sum_{m=1}^{\frac{\ell}{2}}S_{1}^{\Psi^{(2m)}}(\phi,h^{(2m)})\label{l}
\end{eqnarray}
where
\begin{eqnarray}
&&S_{1}^{\Psi^{(2m)}}(\phi,h^{(2m)})=\frac{1}{2m}\int d^{D}z\sqrt{-g}h^{(2m)\mu_{1}...\mu_{2m}}\Psi^{(2m)}_{\mu_{1}...\mu_{2m}}\nonumber\\
&&=\frac{(-1)^{m}}{2m}\int d^{D}z\sqrt{-g}\{h^{(2m)\mu_{1}...\mu_{2m}}\nabla_{\mu_{1}}...\nabla_{\mu_{m}}\phi\nabla_{\mu_{m+1}}...\nabla_{\mu_{2m}}\phi\nonumber\\
&&-\frac{m}{2}h_{\alpha\mu_{m}...\mu_{2m-2}}^{(2m)\alpha\mu_{1}...\mu_{m-1}}\nabla_{(\mu_{1}}...\nabla_{\mu_{m-1}}\nabla_{\mu)}\phi
\nabla^{(\mu_{m}}...\nabla^{\mu_{2m-2}}\nabla^{\mu)}\phi\nonumber\\
&&-\frac{m(D+2m-2)(D+2m-4)}{8L^{2}}
h_{\alpha}^{(2m)\alpha\mu_{1}...\mu_{2m-2}}\nabla_{\mu_{1}}...\nabla_{\mu_{m-1}}\phi
\nabla_{\mu_{m}}...\nabla_{\mu_{2m-2}}\phi
\},\nonumber\\\label{psi2m}
\end{eqnarray}
and the final form of the improved gauge transformations
\begin{eqnarray}
&&\delta_{\epsilon}^{1}\phi(z)=\epsilon^{\mu_{1}\mu_{2}...\mu_{\ell-1}}(z)\nabla_{\mu_{1}}\nabla_{\mu_{2}}...
\nabla_{\mu_{\ell-1}}\phi(z),\label{phigln}\\
&&\delta_{\epsilon}^{0}h^{(2m)\mu_{1}...\mu_{2m}}=2m\nabla^{(\mu_{2m}}\varepsilon^{(2m)\mu_{1}...\mu_{2m-1})},\quad
\delta_{\epsilon}^{0}h_{\alpha}^{(2m)\alpha\mu_{1}...\mu_{2m-2}}=2\varepsilon_{(1)}^{(2m)\mu_{1}...\mu_{2m-2}},\quad\quad\label{tepsm}\\
&&\varepsilon^{(2m)\mu_{1}...\mu_{2m-1}}=\binom{\ell-m-1}{m-1}\epsilon_{(\ell-2m)}^{\mu_{1}...\mu_{2m-1}}.\label{vareps}
\end{eqnarray}
Now we can insert $m=\frac{\ell}{2}$ into (\ref{psil}) and compare our general expression for $S_{1}^{\Psi^{(\ell)}}(\phi,h^{(\ell)})$ with the already known cases of spin two (the energy momentum tensor for the scalar field) (\ref{psi2}) and spin four (\ref{psi4}).
We can easily see that for these cases $S_{1}^{\Psi^{(\ell=2,4)}}(\phi,h^{(\ell)})$ exactly reproduces (\ref{psi2}) and (\ref{psi4}) respectively.
So we found the gauge invariant action for a general spin $l$ gauge field coupled to a scalar and this action has the following property:

\emph{The gauge invariant action $S^{GI}(\phi,h^{(2)},h^{(4)},...,h^{(\ell)})$ for a spin $\ell$ gauge field coupled to a scalar includes gauge invariant actions of the tower of all smaller even spin gauge fields coupled to the same scalar in an analogous way.}

Note that this statement holds true only if we think of an even number of divergencies applied to the gauge parameter as a possible redefinition of gauge parameter of smaller even spin gauge fields, in that case this amazing hierarchy of all smaller even spin currents appear. Another possibility is to regard divergencies of the gauge parameter as gauge transformation for divergencies of the trace of the spin $\ell$ field and make  an appropriate field redefinition. In that case we don't need to introduce smaller spin currents, but the field redefinition will be of another form. The current of spin $\ell$ is the same in both approaches, it is unique, and in the flat space limit reproduces currents constructed in  \cite{Anselmi:1999bb}, \cite{vanDam} and \cite{Petkou} applying a partial integration and field redefinition. The interesting point is that this symmetric form of currents is unique, and the natural generalization of the energy-momentum tensor of the scalar field [\ref{psi2}].

\section{Weyl invariant action for a higher spin field coupled to a scalar}

\quad
In this section we introduce  generalized Weyl transformations for higher spin fields and derive a Weyl invariant action for a higher spin field coupled to a scalar field. Following {\cite{WI}} we write the generalized Weyl transformation for the even spin $l$ field in the form
\begin{eqnarray}
&&\delta_{\sigma}^{0}h^{(\ell)\mu_{1}...\mu_{\ell}}=\ell(\ell-1)\sigma^{(\mu_{1}...\mu_{\ell-2}}g^{\mu_{\ell-1}\mu_{\ell})},\label{hwl}\\
&&\delta_{\sigma}^{0}h_{\alpha}^{(\ell)\alpha\mu_{1}...\mu_{\ell-2}}=2(D+2\ell-4)\sigma^{\mu_{1}...\mu_{\ell-2}},\\
&&\delta_{\sigma}^{1}\phi=\Delta_{\ell}\sigma^{\mu_{1}...\mu_{\ell-2}}\nabla_{\mu_{1}}...\nabla_{\mu_{\ell-2}}\phi.\label{phiwl}
\end{eqnarray}
Then we assume that the Weyl invariant action for a spin $\ell$ field should be accompanied with similar Weyl invariant actions for smaller spin gauge fields and therefore can be constructed from (\ref{l}) adding $\frac{\ell}{2}$ additional terms
\begin{eqnarray}
S^{WI}(\phi,h^{(2)},h^{(4)},...,h^{(\ell)})&=&S^{GI}(\phi,h^{(2)},...,h^{(\ell)})
+\sum_{m=1}^{\ell/2}S_{1}^{r^{(2m)}}(\phi,h^{(2m)}),\label{WIl}
\end{eqnarray}
where each $S_{1}^{r^{(2m)}}$ is gauge invariant itself.
In the case of spin two we had only the linearized Ricci scalar (see (\ref{Sr2})) and for the spin four case we had two terms constructed from the spin four generalization of the Ricci scalar (see (\ref{Sr4})). Now we will see that the generalization of the Ricci scalar for a higher spin field namely the trace of Fronsdal's operator (\ref{frdtr}) (see \cite{Fronsdal},\cite{WI}) is the only gauge invariant combination of two derivatives and a higher spin field which we need to construct the Weyl invariant action (\ref{WIl}) starting from (\ref{l}). We will use the following strategy for solving our problem:
We apply transformation (\ref{hwl})-(\ref{phiwl}) to (\ref{l}) and try to compensate it with the variation of
\begin{eqnarray}
  && \sum_{m=1}^{\ell/2}S_{1}^{r^{(2m)}}(\phi,h^{(2m)}),\ where \nonumber\\
  && S_{1}^{r^{(\ell)}}(\phi,h^{(2)},...,h^{(\ell)})=\nonumber\\
  && =\frac{1}{2}\sum_{m=0}^{\frac{\ell}{2}-1}\xi_{\ell}^{m}\int d^{D}z\sqrt{-g}\nabla_{\mu_{2m+1}}...\nabla_{\mu_{\ell-2}}r^{(\ell)\mu_{1}...\mu_{\ell-2}}\nabla_{\mu_{1}}...\nabla_{\mu_{m}}\phi
  \nabla_{\mu_{m+1}}...\nabla_{\mu_{2m}}\phi \quad\quad\label{Rl}
\end{eqnarray}
introducing necessarily gauge and Weyl transformations for lower spin gauge fields\footnote{Note that (\ref{Rl}) is zero on-shell, when the higher spin gauge field satisfies Fronsdal's free equation of motion, but here the higher spin gauge field is an external off-shell field, and this interaction is a natural generalization of well known nonminimal Weyl invariant coupling with gravity (\ref{1}).}
\begin{eqnarray}
&\delta_{\sigma}h^{(2m)\mu_{1}...\mu_{2m}}=2m(2m-1)C_{\ell}^{m}\sigma_{(\ell-2m)}^{(\mu_{1}...\mu_{2m-2}}g^{\mu_{2m-1}\mu_{2m})},\ \ \ \ m=1,...,\ell/2,\label{wwh}\\
&C_{\ell}^{\ell/2}=1 .\label{wwh1}
\end{eqnarray}
In other words we solve the equation
\begin{eqnarray}
&&\delta^{1}_{\sigma}S^{WI}(\phi,h^{(2)},...,h^{(\ell)})=\delta_{\sigma}^{1}S_{0}+\sum_{s=1}^{\ell/2}\delta_{\sigma}^{0}S_{1}^{\Psi^{(2s)}}
+\sum_{s=1}^{\ell/2}\delta_{\sigma}^{0}S_{1}^{r^{(2s)}}=0 \label{WW}
\end{eqnarray}
which consists of a system of $\ell+1$ equations for $(\ell/2+1)(\ell/2+2)/2$ dependent variables
\footnote{This system includes also (\ref{wwh1}) as an equation for $C_{\ell}^{\ell/2}$.}
 \begin{eqnarray}
   && \triangle_{\ell} ,\label{v1} \\
   && C_{\ell}^{m},\quad m=1,2,\dots,\ell/2, \label{v2}\\
   && \xi_{2s}^{n},\quad n=0,1,\dots s-1;\ s=1,...,\ell/2. \label{v3}
 \end{eqnarray}
but when we find $\xi_{\ell}^{\ell/2-k}$ we also find $\xi_{2s}^{s-k}$ for any $s\geq k$. In other words we find a whole diagonal of this triangle matrix
\begin{eqnarray}
\left(
  \begin{array}{cccccccc}
    C_{\ell}^{1} & C_{\ell}^{2} & . & . & . & C_{\ell}^{\ell/2-1} & C_{\ell}^{\ell/2} & \Delta_{\ell}\\
    \xi_{\ell}^{0} & \xi_{\ell}^{1} & . & . & . & \xi_{\ell}^{\ell/2-2} & \xi_{\ell}^{\ell/2-1} \\
    \xi_{\ell-2}^{0} & \xi_{\ell-2}^{1} & . & . & . & \xi_{\ell-2}^{\ell/2-2} &  \\
    . & . & . & . & . &   \\
    . & . & . & . &   &   \\
    \xi_{4}^{0} & \xi_{4}^{1} &   &   &   &   \\
    \xi_{2}^{0} &   &   &   &   &   \\
  \end{array}\label{matrix}
\right)
\end{eqnarray}
which helps us to solve the whole system. We have two equations for any column of this matrix besides the last, for which we have one equation for
$\Delta$. We start from the last column and go to the left. When we take any column and two equations for that column of variables, we have only two
variables to find if we already solved all columns to the right of that one.\footnote{It is easy to see that the first two rows of (\ref{matrix}) are all we need to find out. The second row  gives the solution for any spin $\ell$. $\xi$-s in lower rows are just particular case and can be determined by putting concrete spin value in a general solution, which means that the independent variables are only first two rows of the (\ref{matrix}) and the number of variables in these two rows is $\ell+1$, just as much as equations we have. This right-to-left method can be used only due to the fact that we solve system for general spin case. This is a  deductive method which we use. Another approach is an inductive method - one could solve equations for concrete cases of spin 2,4,6... and obtaining all rows lower than second (and therefore whole Weyl invariant Lagrangian for lower spins) solve first two rows. Of course this is impossible for general spin $\ell$.}
 That means that our system has a unique solution.
Placing all complicated Weyl variations of (\ref{WW}) into the Appendix C, we present here the resulting system of equations for the unknown variables (\ref{v1})-(\ref{v3}):
\begin{eqnarray}
&&\Delta_{\ell}=1-\frac{D}{2}\label{sys1}\\
&&\frac{(-1)^{\ell/2}}{2}(\Delta_{\ell}-\frac{\ell-2}{2})-(D+2\ell-5)\xi_{\ell}^{\ell/2-1}=0\\
&&(-1)^{m}C_{\ell}^{m}+\sum_{s=m+1}^{\ell/2}mC_{\ell}^{s}\xi_{2s}^{m}=0,\ \ \ (m=1,...,\ell/2-1)\\
&&\frac{(-1)^{m-1}}{2}(m-1)C_{\ell}^{m}-C_{\ell}^{m}(D+4m-5)\xi_{2m}^{m-1}\nonumber\\
&&+\frac{1}{2}\sum_{s=m+1}^{\ell/2}C_{\ell}^{s}[-m(m-1)\xi_{2s}^{m}-(2s-2m+2)(D+2s+2m-5)\xi_{2s}^{m-1}]=0\nonumber\\
&&(m=1,...,\ell/2-1)\label{syslast}
\end{eqnarray}

The solution of this system is universal $\Delta_{\ell}=\Delta=1-\frac{D}{2}$ and
\begin{eqnarray}
\xi_{\ell}^{m}=\frac{(-1)^{m}}{2^{\ell-2m}(\ell/2)}\binom{\ell/2}{m}\frac{(\frac{D}{2}+m-1)_{\ell/2-m}}{(\frac{D+\ell-1}{2}+m-1)_{\ell/2-m}}\\
C_{\ell}^{m}=\frac{(-1)^{\ell/2-m}}{2^{\ell-2m}}\binom{\ell/2-1}{m-1}\frac{(\frac{D}{2}+m-1)_{\ell/2-m}}{(\frac{D-1}{2}+2m)_{\ell/2-m}}.
\end{eqnarray}
These expressions completely fix (\ref{Rl}) and therefore the full Weyl invariant action (\ref{WIl}), and also determine the transformation law for the whole tower of higher spin gauge fields (\ref{wwh}).\footnote{It is easy to see from formula (C.4) that we get also a redefinition of the gauge parameters for all lower even spin fields which in the spin 4 case coincides with formula (\ref{rgp})}

\section*{Conclusion}
We constructed a gauge and generalized Weyl invariant interacting Lagrangian for a linearized higher even spin gauge field and a conformally coupled scalar field in $AdS_{D}$ space. These interactions are unique and nontrivial (nontriviality of the interactions means that they can't be absorbed by a field redefinition from the free action (\ref{S0})).
The resulting Lagrangian for the spin $\ell$ field includes all lower even spin gauge fields also with the same type of interaction with the scalar. These results can be used for constructions of more complicated interactions between different higher spin gauge fields in $AdS$ space (see \cite{MMR2}-\cite{MMR3}).

\bigskip
\section*{Appendix A}
\setcounter{equation}{0}
\renewcommand{\theequation}{A.\arabic{equation}}
\quad Here we present the basic relations between different T-s, M-s and N-s which we use in section 3.
\begin{eqnarray}
&&T(n,k)=(-1)^{m}\sum_{i=0}^{m}\binom{m}{i}T(n-i,k+m-i),\\
&&T(n,k)=(-1)^{m}\sum_{i=0}^{m}\binom{m}{i}T(n-i,k-m),
\end{eqnarray}
and the same for M and N. There is another important relation
\begin{eqnarray}
&&T(n,k)=-M(n-1,k)-\frac{k(k-1)}{2L^{2}}N(n-1,k-1)\nonumber\\
&&-[\frac{(n-k-1)(2D+n-k-4)}{2L^{2}}+\frac{D(D-2)}{4L^{2}}]N(n-1,k)\nonumber\\
&&-\epsilon_{(\ell-n)}^{\mu_{1}...\mu_{n-1}}\nabla_{\mu_{1}}...\nabla_{\mu_{k}}\phi\nabla_{\mu_{k+1}}...\nabla_{\mu_{n-1}}
(\Box-\frac{D(D-2)}{4L^{2}})\phi,
\end{eqnarray}
and the 'symmetrization' relations
\begin{eqnarray}
&&M(2k+1,k)=M(2k+1,k+1)=-\frac{1}{2}M(2k,k),\\
&&N(2k+1,k)=N(2k+1,k+1)=-\frac{1}{2}N(2k,k),\\
&&T(2m,m)=\nabla^{(\alpha}\epsilon_{(\ell-2m)}^{\mu_{1}...\mu_{2m-1})}\nabla_{\mu_{1}}...\nabla_{\mu_{m-1}}\nabla_{\alpha}\phi
\nabla_{\mu_{m}}...\nabla_{\mu_{2m-1}}\phi\nonumber\\
&&+\frac{(m-1)(m-2)}{6L^{2}}N(2m-2,m-1)-\frac{(m-1)(m-2)}{12L^{2}}N(2m-4,m-2) ,\\
&&M(2m-2,m-1)=\epsilon_{(\ell-2m+1)\mu_{m}...\mu_{2m-2}}^{\mu_{1}...\mu_{m-1}}
\nabla_{(\mu_{1}}...\nabla_{\mu_{m-1}}\nabla_{\alpha)}\phi
\nabla^{(\mu_{m}}...\nabla^{\mu_{2m-2}}\nabla^{\alpha)}\phi\nonumber\\
&&+\frac{(m-1)(m-2)}{3L^{2}}N(2m-2,m-1,m-1)-\frac{(m-1)(m-2)}{6L^{2}}N(2m-4,m-2)\nonumber\\
\end{eqnarray}
We must mention here that these relations are satisfied up to full derivatives and therefore admit integration.

\section*{Appendix B}
\setcounter{equation}{0}
\renewcommand{\theequation}{B.\arabic{equation}}
\quad We use the following commutation relations in $AdS_{D}$
\begin{eqnarray}
&&\epsilon^{\mu_{1}\dots\mu_{\ell-1}}[\nabla^{\mu},\nabla_{\mu_{1}}.\dots\nabla_{\mu_{k}}]\phi=
\frac{k(k-1)}{2L^{2}}\epsilon^{\mu\mu_{2}...\mu_{\ell-1}}\nabla_{\mu_{2}}\dots\nabla_{\mu_{k}}\phi,\\
&&[\nabla_{\mu_{1}}\dots\nabla_{\mu_{k}},\nabla^{\mu}]
\epsilon^{\mu_{1}\dots\mu_{\ell-1}}=
\frac{2k(D+\ell-2)-k(k+1)}{2L^{2}}\epsilon_{(k-1)}^{\mu\mu_{k+1}...\mu_{\ell-1}},\\
&&\epsilon^{\mu_{1}\dots\mu_{\ell-1}}[\nabla_{\mu},\nabla_{\mu_{1}}\dots\nabla_{\mu_{k}}]\nabla^{\mu}\phi=
\frac{k(2D+k-3)}{2L^{2}}\epsilon^{\mu_{1}\mu_{2}\dots\mu_{\ell-1}}\nabla_{\mu_{1}}\dots\nabla_{\mu_{k}}\phi,\quad\\
&&\epsilon^{\mu_{1}\dots\mu_{\ell-1}}[\nabla^{2},\nabla_{\mu_{1}}\dots\nabla_{\mu_{k}}]\phi=
\frac{k(D+k-2)}{L^{2}}\epsilon^{\mu_{1}\mu_{2}\dots\mu_{\ell-1}}\nabla_{\mu_{1}}\dots\nabla_{\mu_{k}}\phi,
\end{eqnarray}
where $\epsilon^{\mu_{1}...\mu_{\ell-1}}$ is the symmetric and traceless tensor.
Finally we list all necessary binomial identities
\begin{eqnarray}
&&\sum_{k=0}^{n-m}(-1)^{k}\binom{n}{k}=(-1)^{n-m}\binom{n-1}{m-1},\quad
\sum_{k=0}^{n-m}(-1)^{k}\binom{n}{m+k}=\binom{n-1}{m-1},\quad\quad\\
&&\binom{n}{k}=\binom{n-1}{k-1}+\binom{n-1}{k},\quad
\binom{\ell-m-1}{m-2}=\frac{m-1}{\ell-2m+1}\binom{\ell-m-1}{m-1}.\quad\quad
\end{eqnarray}

\section*{Appendix C}
\setcounter{equation}{0}
\renewcommand{\theequation}{C.\arabic{equation}}
\quad Here we present  all  Weyl variations necessary for the derivation of (\ref{sys1})-(\ref{syslast})
\begin{eqnarray}
&&\delta_{\sigma}^{1}S_{0}=\Delta_{\ell}\int d^{D}z\sqrt{-g}
\{
\sum_{m=1}^{\frac{\ell}{2}-1}\binom{\ell-m-2}{m-1}
\nabla^{(\mu_{2m}}\sigma_{(\ell-2m-1)}^{\mu_{1}...\mu_{2m-1})}\Psi^{(2m)}_{\mu_{1}...\mu_{2m}}\nonumber\\
&&+\sum_{m=1}^{\frac{\ell}{2}-1}\frac{(-1)^{m-1}}{2}\binom{\ell-m-3}{m-1}
\Box\sigma_{(\ell-2m-2)}^{\mu_{1}...\mu_{2m}}\nabla_{\mu_{1}}...\nabla_{\mu_{m}}\phi
\nabla_{\mu_{m+1}}...\nabla_{\mu_{2m}}\phi\nonumber\\
&&+\sum_{m=1}^{\frac{\ell}{2}-1}(-1)^{m}\binom{\ell-m-3}{m-1}
\sigma_{(\ell-2m-2)}^{\mu_{1}...\mu_{2m}}\nabla_{\mu_{1}}...\nabla_{\mu_{m}}\nabla_{\alpha}\phi
\nabla_{\mu_{m+1}}...\nabla_{\mu_{2m}}\nabla^{\alpha}\phi\nonumber\\
&&+O(\frac{1}{L^{2}})\label{swl}
\}.
\end{eqnarray}
We don't have to calculate $O(\frac{1}{L^{2}})$ terms because they can be fixed from flat space considerations and gauge invariance of Fronsdal's operator in AdS.
The first term in (\ref{swl}) can be cancelled by an additional gauge transformation of all gauge fields with spin less than $\ell$.
To cancel other terms we calculate the variation of $\sum_{m=1}^{\ell/2}S_{1}^{\Psi^{(2m)}}(\phi,h^{(2m)})$:
\begin{eqnarray}
&&\delta_{\sigma}^{0}S_{1}^{\Psi^{(2m)}}(\phi,h^{(2)},...,h^{(2m)})\nonumber\\&&
=C_{\ell}^{m}\int d^{D}z\sqrt{-g}
\{-(m-1)[\nabla^{(\mu_{2m-2}}\sigma_{(\ell-2m+1)}^{\mu_{1}...\mu_{2m-3})}\Psi^{(2m-2)}_{\mu_{1}...\mu_{2m-2}}\nonumber\\
&&+\frac{(-1)^{m}}{2}\Box\sigma_{(\ell-2m)}^{\mu_{1}...\mu_{2m-2}}\nabla_{\mu_{1}}...\nabla_{\mu_{m-1}}\phi
\nabla_{\mu_{m}}...\nabla_{\mu_{2m-2}}\phi]\nonumber\\
&&+(-1)^{m}(1-\frac{D}{2})\sigma_{(\ell-2m)}^{\mu_{1}...\mu_{2m-2}}\nabla_{\mu_{1}}...\nabla_{\mu_{m-1}}\nabla_{\alpha}\phi
\nabla_{\mu_{m}}...\nabla_{\mu_{2m-2}}\nabla^{\alpha}\phi\}.
\end{eqnarray}
and  the variation of $\sum_{m=1}^{\ell/2}S_{1}^{r^{(2m)}}(\phi,h^{(2m)})$:
\begin{eqnarray}
&&\delta_{\sigma}^{0}S_{1}^{r^{(\ell)}}=\frac{1}{2}\sum_{m=1}^{\frac{\ell}{2}-1}\int d^{D}z\sqrt{-g}
\{[2m(2m-1)\xi_{\ell}^{m}-2(m-1)(D+4m-8)\xi_{\ell}^{m-1}]\times\nonumber\\
&&\times\nabla^{(\mu_{2m-2}}\sigma_{(\ell-2m+1)}^{\mu_{1}...\mu_{2m-3})}
\Psi^{(2m-2)}_{\mu_{1}...\mu_{2m-2}}\}\nonumber\\
&&-\frac{1}{2}\int d^{D}z\sqrt{-g}\{(\ell-2)(D+2\ell-8)\xi_{\ell}^{\ell/2-1}\nabla^{(\mu_{\ell-2}}
\sigma_{\ell(1)}^{\mu_{1}...\mu_{\ell-3})}\Psi^{(\ell-2)}_{\mu_{1}...\mu_{\ell-2}}
\}\nonumber\\
&&+\frac{1}{2}\sum_{m=1}^{\frac{\ell}{2}-1}\xi_{\ell}^{m}\int d^{D}z\sqrt{-g}\{2m(1-\frac{D}{2})\sigma_{(\ell-2m)}^{\mu_{1}...\mu_{2m-2}}\nabla_{\mu_{1}}...\nabla_{\mu_{m-1}}\nabla_{\alpha}\phi
\nabla_{\mu_{m}}...\nabla_{\mu_{2m-2}}\nabla^{\alpha}\phi
\}\nonumber\\&&+\frac{1}{2}\sum_{m=1}^{\frac{\ell}{2}-1}\int d^{D}z\sqrt{-g}\{[-m(m-1)\xi_{\ell}^{m}-(\ell-2m+2)(D+\ell+2m-5)\xi_{\ell}^{m-1}]\times\nonumber\\
&&\times\Box \sigma_{(\ell-2m)}^{\mu_{1}...\mu_{2m-2}}\nabla_{\mu_{1}}...\nabla_{\mu_{m-1}}\phi
\nabla_{\mu_{m}}...\nabla_{\mu_{2m-2}}\phi\}\nonumber\\
&&-\frac{1}{2}\int d^{D}z\sqrt{-g}\{2(D+2\ell-5)\xi_{\ell}^{\ell/2-1}\Box \sigma^{\mu_{1}...\mu_{\ell-2}}\nabla_{\mu_{1}}...\nabla_{\mu_{\ell/2-1}}\phi
\nabla_{\mu_{\ell/2}}...\nabla_{\mu_{\ell-2}}\phi\}\nonumber\\
&&+O(\frac{1}{L^{2}})
\end{eqnarray}
Then finally we get
\begin{eqnarray}
&&\delta_{\sigma}S^{WI}(\phi,h^{(2)},...,h^{(\ell)})=
\delta_{\sigma}^{1}S_{0}+\sum_{s=1}^{\ell/2}\delta_{\sigma}^{0}S_{1}^{\Psi^{(2s)}}
+\sum_{s=1}^{\ell/2}\delta_{\sigma}^{0}S_{1}^{r^{(\ell)}}\nonumber\\
&&=\sum_{m=1}^{\ell/2-1}\int d^{D}z\sqrt{-g}
\{\binom{\ell-m-2}{m-1}\Delta_{\ell}-mC_{\ell}^{m+1}[1+(D+4m-4)\xi_{2m+2}^{m}]\nonumber\\
&&+\frac{1}{2}\sum_{s=m+2}^{\ell/2}[(2m+2)(2m+1)\xi_{2s}^{m+1}-2m(D+4m-4)\xi_{2s}^{m}]\}\times\nonumber\\
&&\times\nabla^{(\mu_{2m}}\sigma_{(\ell-2m-1)}^{\mu_{1}...\mu_{2m-1})}\Psi^{(2m)}_{\mu_{1}...\mu_{2m}}\nonumber\\
&&+\int d^{D}z\sqrt{-g}\{\frac{(-1)^{\ell/2}}{2}(\Delta_{\ell}-\frac{\ell-2}{2})-(D+2\ell-5)\xi_{\ell}^{\ell/2-1}\}\times\nonumber\\
&&\times\Box \sigma^{\mu_{1}...\mu_{\ell-2}}\nabla_{\mu_{1}}...\nabla_{\mu_{\ell/2-1}}\phi\nabla_{\mu_{\ell/2}}...\nabla_{\mu_{\ell-2}}\phi\nonumber\\
&&+\sum_{m=1}^{\ell/2-1}\int d^{D}z\sqrt{-g}\{\frac{(-1)^{m}}{2}[\binom{\ell-m-2}{m-2}\Delta_{\ell}-(m-1)C_{\ell}^{m}]-C_{\ell}^{m}(D+4m-5)\xi_{2m}^{m-1}\nonumber\\
&&+\frac{1}{2}\sum_{s=m+1}^{\ell/2}C_{\ell}^{s}[-m(m-1)\xi_{2s}^{m}-(2s-2m+2)(D+2s+2m-5)\xi_{2s}^{m-1}]\}\times\nonumber\\
&&\times\Box\sigma_{(\ell-2m)}^{\mu_{1}...\mu_{2m-2}}\nabla_{\mu_{1}}...\nabla_{\mu_{m-1}}\phi
\nabla_{\mu_{m}}...\nabla_{\mu_{2m-2}}\phi\nonumber\\
&&+\int d^{D}z\sqrt{-g}\{(-1)^{\ell/2}(1-\frac{D}{2}-\Delta_{\ell})\sigma^{\mu_{1}...\mu_{\ell-2}}
\nabla_{\mu_{1}}...\nabla_{\mu_{\ell/2-1}}\nabla_{\alpha}\phi
\nabla_{\mu_{\ell/2}}...\nabla_{\mu_{\ell-2}}\nabla^{\alpha}\phi\}\nonumber\\
&&+\sum_{m=1}^{\ell/2-1}\int d^{D}z\sqrt{-g}\{(-1)^{m-1}\binom{\ell-m-2}{m-2}\Delta_{\ell}+(-1)^{m}(1-\frac{D}{2})C_{\ell}^{m}\nonumber\\
&&+(1-\frac{D}{2})\sum_{s=m+1}^{\ell/2}mC_{\ell}^{s}\xi_{2s}^{m}\}\sigma_{(\ell-2m)}^{\mu_{1}...\mu_{2m-2}}
\nabla_{\mu_{1}}...\nabla_{\mu_{m-1}}\nabla_{\alpha}\phi\nabla_{\mu_{m}}...\nabla_{\mu_{2m-2}}\nabla^{\alpha}\phi
\end{eqnarray}
From this expression we can derive our system of equations (\ref{sys1})-(\ref{syslast}).

\subsection*{Acknowledgements}
This work is supported in part by Alexander von Humboldt Foundation under 3.4-Fokoop-ARM/1059429 and ANSEF 2009.
Work of K.M. made with partial support of CRDF-NFSAT UCEP06/07 and CRDF-NFSAT-SCS MES RA ECSP 09\_01/A-31.

\end{document}